\documentclass{article}
\usepackage{amsmath}
\usepackage{graphicx}
\usepackage{epstopdf}
\usepackage{amssymb}
\usepackage{float}

\begin{document}

\begin{center}\LARGE{Quantum Incompressibility of a Falling Rydberg Atom, and a
Gravitationally-Induced Charge Separation Effect in Superconducting Systems}\\ \vspace{.2 in}\large{R. Y. Chiao\footnote{Schools of Natural Sciences and Engineering, University of California, P. O. Box 2039, Merced, CA 95344, USA, corresponding author, e-mail: rchiao@ucmerced.edu\\}, S. J. Minter\footnote{School of Natural Sciences, University of California, 5200 N. Lake Rd., Merced, CA  95343, USA\\}, K. Wegter-McNelly\footnote{Boston University, School of Theology, 745 Commonwealth Ave., Boston, MA  02215, USA\\}, L. A. Martinez$^2$\\ \vspace{.2in}
 \vspace{.1in}
2010,`2nd Vienna Symposium for the Foundations of Modern
Physics' Festschrift MS for \textit{Foundations of Physics}}\end{center}

\begin{abstract}Freely falling point-like objects converge toward the center of the Earth.  Hence the gravitational field of the Earth is inhomogeneous, and possesses a tidal component.  The free fall of an extended quantum mechanical object such as a hydrogen atom prepared in a high principal-quantum-number state, i.e. a circular Rydberg atom, is predicted to fall more slowly than a classical point-like object, when both objects are dropped from the same height above the Earth's surface.  This indicates that, apart from transitions between quantum states, the atom exhibits a kind of quantum mechanical incompressibility during free fall in inhomogeneous, tidal gravitational fields like those of the Earth.

A superconducting ring-like system with a persistent current circulating around it behaves like the circular Rydberg atom during free fall.  Like the electronic wavefunction of the freely falling atom, the Cooper-pair wavefunction is quantum mechanically incompressible.  The ions in the lattice of the superconductor, however, are not incompressible, since they do not possess a globally coherent quantum phase.  The resulting difference during free fall in the response of the \emph{nonlocalizable} Cooper pairs of electrons and the \emph{localizable} ions to inhomogeneous gravitational fields is predicted to lead to a charge separation effect, which in turn leads to a large Coulomb force that opposes the convergence caused by the tidal gravitational force on the superconducting system.

A \textquotedblleft Cavendish-like\textquotedblright\ experiment is proposed for observing the charge separation effect induced by inhomogeneous gravitational fields in a superconducting circuit.  The charge separation effect is determined to be limited by a pair-breaking process that occurs when low frequency gravitational perturbations are present. \vspace*{2.7in}
\end{abstract}

\vspace{3in}

\pagebreak

\section{Introduction}

\begin{figure}[tbp]\begin{center}
\label{radial-converging-geodesics} 
\includegraphics[width=2in]{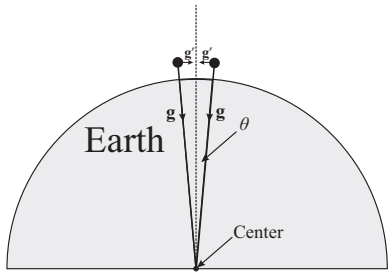}\end{center}
\caption{Two nearby, freely falling, point-like objects dropped from the
same height above the Earth's surface follow converging trajectories that
are inclined at a slight angle $\protect\theta $ with respect to the
vertical plumb line equidistant between them. According to a distant
inertial observer, the radial convergence of these objects' trajectories
towards the center of the Earth causes them to undergo small horizontal
components of acceleration $\mathbf{g}^{\prime }$ of the radial acceleration 
$\mathbf{g}$. These components are equivalent to a tidal gravitational force
that, in a Newtonian picture, causes the two objects to converge toward one
another.}
\end{figure}

Experiments at the frontiers of quantum mechanics and gravity are rare. We
would like to explore in this essay, in honor of Danny Greenberger and Helmut
Rauch, situations which could lead to such experiments. The key is to
understand the phenomenon of \textquotedblleft quantum
incompressibility\textquotedblright\ of macroscopically coherent quantum
matter in the presence of inhomogeneous, tidal gravitational fields, such as
the Earth's. See Figure \ref{radial-converging-geodesics}.

As an example of \textquotedblleft quantum
incompressibility\textquotedblright\ during free fall of an extended quantum
object,\ let us first consider the single electron of a circular Rydberg
atom \cite{Hulet-and-Kleppner} (ignoring electron spin), which is prepared
in the state 
\begin{equation}
\left\vert n,\text{ }l=n-1,\text{ }m=n-1\right\rangle \text{ ,}
\label{ring-like state}
\end{equation}%
where $n$ is the principal quantum number, which is large, i.e., $n\gg 1$,
and $l=n-1$ is the maximum possible orbital angular momentum quantum number
for a given $n$, and $m=l=n-1$ is the maximum possible azimuthal quantum
number for a given $l$, i.e., the \textquotedblleft
stretched\textquotedblright\ state. The $z$ axis has been chosen to be the
local vertical axis located at the center of mass of the atom. Then the
wavefunction of this electron in polar coordinates $(r,\theta ,\phi )$ of
the hydrogenic atom in this state is given by \cite[p. 253]%
{Haroche-and-Raimond}%
\begin{equation}
\Psi _{n,n-1,n-1}(r,\theta ,\phi )=N_{n,n-1,n-1}(r\sin \theta e^{i\phi
})^{n-1}\exp \left( -\frac{r}{na_{0}}\right) \text{ ,}  \label{ring-like wf}
\end{equation}%
where $N_{n,n-1,n-1}$ is a normalization constant. 

\begin{figure}[tbp]\begin{center}
\label{Rydberg} \includegraphics[width=2in]{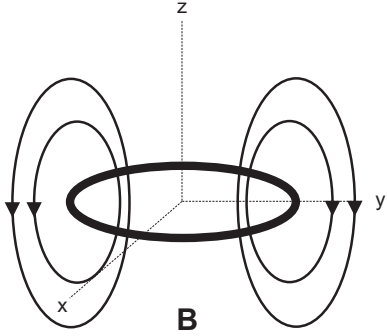}\end{center}
\caption{A circular Rydberg atom in the state $\left\vert n,\text{ }l=n-1,%
\text{ }m=n-1\right\rangle $ has a strongly peaked, ring-like probability
distribution, i.e., an \textquotedblleft electron cloud,\textquotedblright\
indicated by the heavy black loop. Currents in this state lead to a magnetic
field $\mathbf{B}$, indicated by the directional loops. The $z$ axis is the
local vertical axis.}
\end{figure}

The probability density
associated with this wavefunction has the form of a strongly peaked
distribution which lies on the horizontal $(x,y)$ plane, in the shape of a
ring of radius 
\begin{equation}
a_{n}=n^{2}\frac{\hbar ^{2}}{me^{2}}=n^{2}a_{0}\text{ ,}
\label{size of Rydberg atom}
\end{equation}%
where $a_{0}$ is the Bohr radius. Thus one recovers the Bohr model of the
hydrogen atom in the correspondence-principle limit of large $n$. This
ring-like probability distribution is illustrated in Figure 2.
\qquad

The question we would like to address here is this: How does the size of this
atom change with time as it undergoes free fall in Earth's inhomogeneous,
tidal gravitational field?

\section{An analogy}

The magnetic moment of the Rydberg atom in the state (\ref{ring-like wf}) is
quantized, and is given by%
\begin{equation}
\mu _{n}=n\frac{e\hbar }{2m}=n\mu _{B}\text{ ,}
\end{equation}%
where $\mu _{B}$ is the Bohr magneton and $n$ is an integer.

The electron current density in the ring-like structure of a circular
Rydberg atom in Figure 2 is similar to that of a persistent
supercurrent of Cooper pairs in a superconducting ring with a quantized flux
given by%
\begin{equation}
\Phi _{n}=n\Phi _{0}=n\frac{h}{2e}\text{ ,}
\end{equation}%
where $\Phi _{0}$ is the flux quantum and $n$ is an integer. The quantum
incompressibility of the ring-like structure of a circular Rydberg atom, and
the quantum incompressibility of the Cooper pairs of electrons in a
superconducting ring, both arise from the same quantum mechanical principle,
namely, the single-valuedness of the wavefunction after one round trip
around the ring, which follows from the condition%
\begin{equation}
\oint\limits_{\text{ring}}\nabla \varphi \cdot d\mathbf{l}=\Delta \varphi
=2\pi m\text{ ,}
\end{equation}%
where $\varphi $ is the phase of the wavefunction, and $m$ is an integer
corresponding to the state under consideration. Another necessary condition
for quantum incompressibility is the existence of a substantial energy gap
separating the $m$th state from adjacent states of the system.

The analogy between the Rydberg atom and the superconducting ring is not a
perfect one, since the selection rules for allowed transitions between
adjacent states will be different in the two cases. The transitions $%
n\rightarrow n-1$ and $n\rightarrow n+1$ are electric-dipole allowed for the
Rydberg atom, whereas the transitions $n\rightarrow n-1$ and $n\rightarrow
n+1$ between adjacent flux-trapping states of the superconducting ring are
highly forbidden. This is because a macroscopic number of identical Cooper
pairs of electrons must all simultaneously jump from a state with $n\hbar $
units to a state with $\left( n-1\right) \hbar $ units or with $\left(
n+1\right) \hbar $ units of angular momentum per electron pair. Hence the
persistent current of a superconducting ring is highly metastable, and does
not change with time, unless a macroscopic quantum
transition occurs.

If the characteristic frequency of an external perturbation, such as that of
the tidal gravitational fields acting on the system during free fall in
Earth's gravity, is much less than the smallest energy gap of an allowed
transition divided by Planck's constant, then the system cannot make a
transition (i.e., a \textquotedblleft quantum jump\textquotedblright ) out
of its initial state. Thus it must stay rigidly in its initial state. (For a
Rydberg atom with $n\simeq 100$, this transition frequency lies in the
gigahertz range, so that this assumption is well satisfied.) The size of the
circular Rydberg atom and the size of the persistent currents of the
superconducting ring will therefore remain constant in time during
perturbations arising from Earth's tidal fields during free fall, apart from
a sequence of possible \textquotedblleft quantum jumps\textquotedblright\ in
a \textquotedblleft quantum staircase,\textquotedblright\ though such
transitions occur only in highly unusual circumstances in the gravitational
field of the Earth.

\section{The Quantum Incompressibility of the Rydberg Atom}

It has been previously shown \cite{Parker} that the electron wavefunction of
a hydrogen atom will be altered by the presence of curvature. \ The
Hamiltonian operator for such an operator is given by%
\begin{equation}
H=H_{0}+H_{1},
\end{equation}%
where $H_{0}$ is the unperturbed operator, and $H_{1}$ is an interaction
Hamiltonian due to curvature. \ This interaction operator was shown to be%
\begin{equation}
H_{P}\equiv H_{1}=\frac{1}{2}m_{e}R_{0l0m}x^{l}x^{m},  \label{H_I}
\end{equation}%
where $m_{e}$ is the mass of the electron, $R_{0l0m}$ are components of the
Riemann curvature tensor, and $x^{l}$ and $x^{m}$ are components of the
position operator of the electron. Equation (\ref{H_I}) is obtained by
taking the nonrelativistic limit of the Dirac equation embedded into curved
spacetime \cite{Parker PRL}\cite{Parker PRD1980}. \ The subscript $P$ is
used for Parker, the author of the papers referenced here, and his index
notation is adopted as well. \ Superscripts $l$ and $m$ are contravariant
indices satisfying $l,m=1,2,3$, and should not be confused with quantum
number $l$ or mass $m$. \ The index $0$ refers to the time components. \
Greek indices are spacetime indices, and Latin indices are space indices
only. \ The Einstein summation convention is used here as well, in which
repeated indices are summed over all possible values.

Expressing these components of the Riemann tensor in terms of the
gravitational scalar potential, which satisfies%
\begin{equation}
-\nabla \Phi =\mathbf{g},
\end{equation}%
one can show that near the surface of the earth, the interaction Hamiltonian
is given by 
\begin{equation}
H_{P}=\frac{m_{e}g}{2R_{E}}\left( x^{2}+y^{2}-2z^{2}\right)
\end{equation}%
in Cartesian coordinates, or%
\begin{equation}
H_{P}=\frac{m_{e}g}{2R_{\text{E}}}\left[ r^{2}\left( 3\sin ^{2}\theta
-2\right) \right]
\end{equation}%
in spherical coordinates. \ It is worth noting that this interaction
Hamiltonian is proportional to the second-degree, zero-order spherical
harmonic $Y_{2}^{0}.$ \ Thus, this interaction Hamiltonian represents a
rank-2 angular momentum operator, since its angular dependence is
quadrupolar \cite{Gill}. \ One immediate consequence is that, by the
Wigner-Eckart theorem, there will be no effect on a hydrogen atom in a state
with zero angular momentum (i.e. where $l=0$), to first order, through this
interaction Hamiltonian.

The energy shift associated with $H_{P}$ is given by%
\begin{equation}
\Delta E_{P}=\frac{\left\langle \Psi |H_{P}|\Psi \right\rangle }{%
\left\langle \Psi |\Psi \right\rangle }.
\end{equation}%
Taking the expectation values with respect to the stretched-state
wavefunction in (\ref{ring-like wf}), one obtains%
\begin{equation}
\Delta E_{P}\approx \frac{m_{e}g}{2R_{\text{E}}}a_{n}^{2}=\frac{%
m_{e}ga_{0}^{2}}{2R_{\text{E}}}n^{4},
\end{equation}%
where the approximation is valid for large values of principal quantum
number $n.$

In addition to the energy shift derived from the interaction Hamiltonian $%
H_{P}$, one can use DeWitt's minimal coupling rule, which is associated with
an asymptotically flat coordinate system, unlike the operator $H_{P}$, which
is associated with the transverse-traceless gauge often used in general
relativity. \ Let us show that quantum incompressibility is predicted to
occur in a circular Rydberg atom, starting from DeWitt's minimal coupling
rule. The DeWitt Hamiltonian for a freely falling hydrogenic atom, such as a
circular Rydberg atom in presence of weak electromagnetic and gravitational
fields, is given in SI units by%
\begin{equation}
H=\frac{1}{2m}\left( \mathbf{p}-e\mathbf{A}-m\mathbf{h}\right) ^{2}+\frac{%
e^{2}}{4\pi \varepsilon _{0}r}\text{ ,}
\label{DeWitt Hamiltonian for Rydberg atom}
\end{equation}%
where $\mathbf{A}$ is the electromagnetic vector potential, and $\mathbf{h}$%
\ is DeWitt's gravitational vector potential \cite{DeWitt}.

Before perfoming any quantitative analyses, it is stated here beforehand
that while the Hamiltonian in (\ref{DeWitt Hamiltonian for Rydberg atom})
applies to the entire atom, one can apply a center-of-mass and relative
coordinates transformation to the unperturbed Hamiltonian, treating the
interaction Hamiltonians containing the vector potentials as perturbations.
\ The fact that the proton has a much larger mass and is located at a much
smaller distance from the center of mass causes the proton contributions to
the energy shifts to be negligible compared to those of the electron
contributions. \ 

The interaction Hamiltonian for the $\mathbf{A\cdot A}$ term (the
\textquotedblleft Landau diamagnetism term\textquotedblright\ \cite%
{Landau&Lifshitz}) is given by\ \cite{A.p}%
\begin{equation}
H_{\mathbf{A\cdot A}}=\frac{e^{2}}{2m_{e}}\mathbf{A\cdot A}\text{ .}
\end{equation}%
In the symmetric gauge, where $\mathbf{A}=\frac{1}{2}\mathbf{B}\times 
\mathbf{r=-}\frac{1}{2}B(y\mathbf{e}_{x}-x\mathbf{e}_{y})$, for $\mathbf{B}=B%
\mathbf{e}_{z}$, and where $\mathbf{e}_{x},\mathbf{e}_{y},$ and $\mathbf{e}%
_{z}$ are the unit vectors along the $x$, $y$, and $z$ axes, respectively,
this yields%
\begin{equation}
H_{\mathbf{A\cdot A}}=\frac{e^{2}B^{2}}{8m_{e}}(x^{2}+y^{2})\text{,}
\end{equation}%
in Cartesian coordinates, or%
\begin{equation}
H_{\mathbf{A\cdot A}}=\frac{e^{2}B^{2}}{8m_{e}}r^{2}\sin ^{2}\theta 
\end{equation}%
in spherical coordinates, where $\theta $ is the azimuthal angle. \ The
energy shift in first-order perturbation theory resulting from the presence
of the $\mathbf{A}$ field is given by%
\begin{equation}
\Delta E_{\mathbf{A\cdot A}}=\frac{e^{2}B^{2}}{8m_{e}}\left\langle \Psi
_{nlm}\left\vert r^{2}\sin ^{2}\theta \right\vert \Psi _{nlm}\right\rangle 
\text{ .}  \label{energy A}
\end{equation}%
Recalling that the wavefunction for the circular Rydberg state is given by (%
\ref{ring-like wf}), the expectation value in (\ref{energy A}) becomes%
\begin{equation}
\frac{\left\langle \Psi \left\vert r^{2}\sin ^{2}\theta \right\vert \Psi
\right\rangle }{\left\langle \Psi |\Psi \right\rangle }\approx
(n^{2}a_{0})^{2}=a_{n}^{2}\text{ }  \label{mean-square-size}
\end{equation}%
for large values of the principal quantum number $n$, where $n>>1$. It
follows that the first-order energy shift of the atom in the presence of a
magnetic field is%
\begin{equation}
\Delta E_{\mathbf{A\cdot A}}\approx \frac{e^{2}a_{n}^{2}}{8m_{e}}\text{ }%
B^{2}\text{.}  \label{A.A_energy_shift}
\end{equation}%
This result implies that, in first-order perturbation theory, the size of
the atom does not change in the presence of the applied DC magnetic field,
in the sense that the root-mean-square transverse size of the atom, which is
given by%
\begin{equation}
\left. a_{n}\right\vert _{\text{rms}}=\sqrt{\left\langle \Psi \left\vert
r^{2}\sin ^{2}\theta \right\vert \Psi \right\rangle }=a_{n}
\end{equation}%
does not change with time during the application of the DC magnetic field.
Moreover, the wavefunction $\Psi _{n,n-1,n-1}$ remains unaltered in
first-order perturbation theory in the presence of a weak applied field.
Furthermore, this is still true for applied magnetic fields which vary
sufficiently slowly in time, so that no transitions (i.e., \textquotedblleft
quantum jumps\textquotedblright ) can occur out of the initial state of the
system $\Psi _{n,n-1,n-1}$. The concept of the \textquotedblleft quantum
incompressibility\textquotedblright\ of a Rydberg atom is thus a valid
concept during the application of sufficiently weak, and sufficiently slowly
varying, magnetic fields.

The energy shift given by (\ref{A.A_energy_shift}) causes the atom to become
a \emph{low-field seeker} in inhomogeneous magnetic fields through the
relationship%
\begin{equation}
\left( \mathbf{F}_{_{\mathbf{A\cdot A}}}\right) _{n}=-\nabla \left( \Delta
E_{\mathbf{A\cdot A}}\right) _{n}\approx -\frac{e^{2}a_{n}^{2}}{8m_{e}}\text{
}\nabla \left( B^{2}\right) \text{ ,}  \label{F_A.A}
\end{equation}%
where $\left( \mathbf{F}_{_{\mathbf{A\cdot A}}}\right) _{n}$ is the force on
the atom in the ring-like state (\ref{ring-like wf}) in the presence of an
inhomogeneous magnetic field.

Next, let us consider the more interesting case of when weak tidal
gravitational fields are present without any accompanying electromagnetic
fields, i.e., when $\mathbf{h}\neq \mathbf{0}$ and $\mathbf{A}=\mathbf{0}$.
As before, the atom is initially prepared in the state given by (\ref%
{ring-like wf}) before it is released into free fall in the Earth's
inhomogeneous gravitational field. The $z$ axis, which goes through the
center of mass of the atom, is chosen to be the local vertical axis of the
Earth's field. The horizontal tidal gravitational fields of the Earth
experienced during free fall by the atom, as observed in the coordinate
system of a distant inertial observer, where the $(x,y)$ plane is the local
horizontal plane, will be given by%
\begin{equation}
\mathbf{h}(x,y,t)=\mathbf{v}(x,y,t)=\mathbf{g}^{\prime }t=\frac{gt}{R_{E}}(%
\mathbf{e}_{x}x+\mathbf{e}_{y}y)\text{ ,}  \label{tidal h field}
\end{equation}%
where $\mathbf{v}(x,y,t)$ is the velocity of a freely falling, point-like
test particle located at $(x,y)$ and observed at time $t$ by the distant
inertial observer \cite{Prague}, $\mathbf{g}^{\prime }$ is the horizontal
component of Earth's gravitational acceleration arising from the radial
convergence of free-fall trajectories towards the center of the Earth as
seen by this observer (see Figure \ref{radial-converging-geodesics}), $R_{E}$
is the radius of the Earth, and $\mathbf{e}_{x}$ and $\mathbf{e}_{y}$ are
respectively the unit vectors pointing along the $x$ and the $y$ axes, in
this observer's coordinate system. In (\ref{tidal h field}) we have assumed
that the horizontal excursions of the electron in $x$ and $y$ are very small
compared to the Earth's radius. The interaction Hamiltonian for the $\mathbf{%
h}\cdot \mathbf{h}$ term in (\ref{DeWitt Hamiltonian for Rydberg atom}) is
given by%
\begin{equation}
H_{\mathbf{h}\cdot \mathbf{h}}=\frac{m}{2}\mathbf{h}\cdot \mathbf{h}=\frac{%
m_{e}g^{2}t^{2}}{2R_{E}^{2}}(r^{2}\sin ^{2}\theta )\text{ .}
\end{equation}%
Therefore, the shift in energy of the atom in the circular Rydberg state,
due to the Earth's tidal fields given by (\ref{tidal h field}), is given in
first-order perturbation theory by%
\begin{equation}
\Delta E_{\mathbf{h}\cdot \mathbf{h}}=\frac{m_{e}g^{2}t^{2}}{2R_{E}^{2}}%
\frac{\left\langle \Psi \right\vert r^{2}\sin ^{2}\theta \left\vert \Psi
\right\rangle }{\left\langle \Psi |\Psi \right\rangle }\approx \frac{%
m_{e}a_{n}^{2}}{2R_{E}^{2}}g^{2}t^{2}  \label{energy shift for h}
\end{equation}%
for large values of $n$, where $n>>1$. Once again, since the expectation
value in (\ref{energy shift for h}) is the mean-square transverse size of
the atom, this implies that the size of the atom does not change during free
fall, according to first-order perturbation theory. In other words, the atom
is \textquotedblleft quantum incompressible\textquotedblright\ in the
presence of the inhomogeneous, tidal fields of the Earth, just like in the
case of the atom in the presence of an applied DC magnetic field, as long as
transitions (i.e., \textquotedblleft quantum jumps\textquotedblright ) out
of the initial quantum state $\Psi _{n,n-1,n-1}$\ cannot occur. This
conclusion is valid assuming that the characteristic frequency of the
applied tidal fields is much less than the gap frequency (i.e., the energy
gap divided by Planck's constant, which is typically on the order of
gigahertz for $n\sim 100$) corresponding to a quantum transition from the $n$%
th state to the nearest adjacent allowed states, and assuming that the tidal
gravitational field of the Earth is sufficiently weak.

In the gravitational case, just as in the magnetic case, the energy-level
shift caused by the tidal perturbations arising from the Earth's
inhomogeneous gravitational field, leads to a force on the atom. This force
causes the atom to become a \emph{low-field seeker} in the inhomogeneous
gravitational\emph{\ }field of the Earth through the relationship%
\begin{equation}
\left( \mathbf{F}_{_{\mathbf{h\cdot h}}}\right) _{n}=-\nabla \left( \Delta
E_{\mathbf{h\cdot h}}\right) _{n}\approx -\frac{1}{2}m_{e}a_{n}^{2}t^{2}%
\text{ }\nabla \left( \frac{g^{2}}{R_{E}^{2}}\right) \text{ .}  \label{F_h.h}
\end{equation}%
Thus a hydrogen atom in a circular Rydberg state, which is an \emph{extended}
quantum object, will fall slightly more slowly than a point-like classical
test particle which is simultaneously released into free fall along with the
atom near the center of mass of the atom in Earth's inhomogeneous field.

The gravitational, Landau-like energy shifts of the atom given by (\ref%
{energy shift for h}) are much too small to be measured directly in
Earth-bound experiments using current technology, but in principle they can
be measured spectroscopically by monitoring the frequencies of transitions
between adjacent Rydberg states, for example, in a satellite laboratory
which is in a highly elliptical orbit around the Earth. It is therefore a
genuine physical effect.

One might be tempted to say that the force given by (\ref{F_h.h}) questions
the universal applicability of the equivalence principle, i.e., the
universality of free fall. But it must be kept in mind that the equivalence
principle applies strictly to point objects only, inside of
which tidal effects can be neglected. This is manifestly not the case for
the extended quantum systems being considered here.

\section{A superconducting circuit consisting of two cubes joined coherently
by two parallel wires}

\begin{figure}[tbp]\begin{center}
\label{2-sc-cubes-with-2-sc-wires} 
\includegraphics[width=3in]{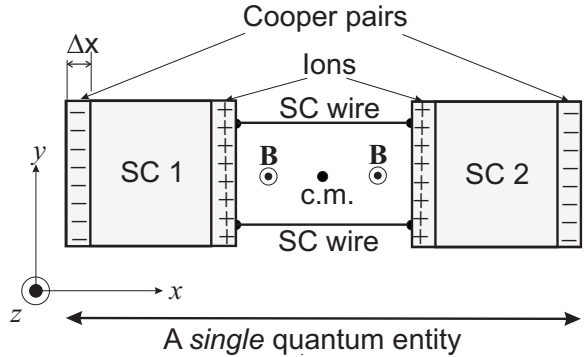}\end{center}
\caption{Two superconducting cubes, SC 1 and SC 2, which are undergoing free
fall in Earth's inhomogeneous gravitational field, are connected by means of
two thin superconducting wires, which establishes \emph{quantum coherence}
throughout the system, and makes it a single quantum entity with a center of
mass (\textquotedblleft c.m.\textquotedblright ) located in the middle.\ A
persistent current through the wires traps a $\mathbf{B}$ field inside this
superconducting circuit, much like in the circular Rydberg atom. All
dimensions of the cubes and the length of the wire are given by the same
distance $L$. The $z$ axis denotes the local vertical axis passing through
\textquotedblleft c.m.\textquotedblright .}
\end{figure}

The analogy between the Rydberg atom and superconducting ring suggests a
simple experiment to test the idea of \textquotedblleft quantum
incompressibility\textquotedblright\ during free fall, which can be
performed in an ordinary laboratory. Consider a horizontal system consisting
of two superconducting cubes joined by two parallel superconducting wires to
form a superconducting circuit. See Figure 3.

When a coherent quantum connection between the two cubes is not present
(due, say, to the effect of heating coils wrapped around the midsections of
both wires which drive them to a normal state by heating them above their transition
temperature, so that the coherent quantum connection between the cubes is
thereby destroyed), the centers of masses of the two spatially separated
cubes, which will have decohered with respect to each other, will follow the
converging free-fall trajectories shown in Figure \ref%
{radial-converging-geodesics}, which are inclined at a slight angle $\theta $
with respect to the vertical plumb line passing through the midpoint
\textquotedblleft c.m.\textquotedblright , with%
\begin{equation}
\theta \approx \frac{L}{R_{E}}\text{ ,}
\end{equation}%
where $L$ is separation of the two cubes, which is also their dimensions
(thus chosen for simplicity), and $R_{E}$ is the radius of the Earth. It
should be noted that the \emph{decoherence}, and therefore the \emph{spatial
separability}, of entangled states arising from perturbations due to the
environment \cite{Zurek}, is a necessary precondition for the applicability
of the equivalence principle \cite{Prague}, so that here the universality of
free fall can be applied to the free-fall trajectories of the \emph{%
disconnected} superconducting cubes \cite{geodesics}.

When a coherent connection is\emph{\ }present between the two cubes, they
will become a single, macroscopic quantum object like the freely-falling
Rydberg atom. The Cooper pairs of electrons of the system will then remain
motionless with respect to the midpoint \textquotedblleft
c.m.\textquotedblright , since their macroscopic wavefunction corresponds to
a zero-momentum eigenstate relative to c.m., and therefore, by the
uncertainty principle, the electrons are completely nonlocalizable within
the entire, coherently connected two-cube system. The Cooper pairs of
electrons, like the electron in Rydberg atom, will then exhibit quantum
incompressibility\ during free fall. This follows from the fact that the
mean-squared size of the coherent electrons of the \emph{entire}
superconducting system remains unchanged in response to the tidal
gravitational fields of the Earth, according to first-order perturbation
theory.

However, the ions, which have undergone decoherence due to the environment 
\cite{Zurek}\cite{Prague}, are completely \emph{localizable}, and therefore,
by the equivalence principle, will want to follow the free-fall trajectories
that converge onto the center of the Earth shown in Figure \ref%
{radial-converging-geodesics}. By contrast, the Cooper pairs of electrons
will remain coherent during free fall, since they are protected from
decoherence by the BCS energy gap \cite{Prague}, and will therefore remain
completely \emph{nonlocalizable}, since they will remain in a zero-momentum
eigenstate. This difference in the motion of the ions and of the Cooper
pairs of electrons will then lead to the charge-separation effect indicated
in Figure 3, in which the ions will be
extruded through the innermost faces of the cubes, because of the
convergence of their radial trajectories that point towards the center of
the Earth, and in which the Cooper pairs of electrons, which resist this
convergence, will be extruded through the outermost faces of the cubes.

\begin{figure}[tbp]\begin{center}
\label{Ball-and-stick-model} 
\includegraphics[width=3in]{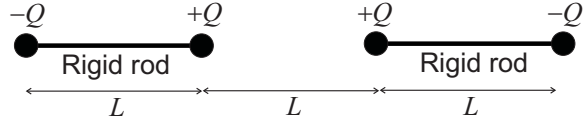}\end{center}
\caption{Ball with charges $-Q$ and $+Q$ are attached to rigid rods with
lengths $L$ to form two dumbbells, which model the configuration of charges
in Figure 3. The two innermost
charges, both of which are $+Q$, are separated by a distance $L$. These two
innermost charges dominate the Coulomb force between the two dumbbells, so
that the net force is a repulsive one.}
\end{figure}

The Cooper pairs of electrons in the zero-momentum eigenstate,
which remain at rest rigidly with respect to the \emph{global} c.m.\ of the
entire, coherent two-cube system, will therefore be displaced away from the ions by a
distance $\Delta x$ on left face of the left cube, and also on the right
face of the right cube. The resulting charge configuration can be
approximated by a ball-and-stick model of two charged dumbbells shown in
Figure \ref{Ball-and-stick-model}.

On the one hand, the net Coulomb force between the two dumbbells in Figure %
\ref{Ball-and-stick-model} is given by%
\begin{equation}
F_{\text{Coulomb}}=\alpha \frac{Q^{2}}{4\pi \varepsilon _{0}L^{2}}\text{ ,}
\label{alpha}
\end{equation}%
where $\alpha $ is a pure number on the order of unity (this follows from
dimensional considerations, since $L$ is the only distance scale in the
problem).

On the other hand, the tidal gravitational force between the cubes\ in
Figure 3 is given by%
\begin{equation}
F_{\text{Tidal}}=Mg^{\prime }\text{ ,}
\end{equation}%
where $M$ is the mass of the cube (which is mainly due to the ions), and%
\begin{equation}
g^{\prime }=g\sin \theta \approx g\tan \theta \approx g\theta \approx
gL/R_{E}
\end{equation}%
is the horizontal component of the acceleration due to Earth's gravity
acting on the centers of the cubes, which is directed towards the midpoint
c.m.\ of the two cubes. Thus, in equilibrium,%
\begin{equation}
F_{\text{Coulomb}}=F_{\text{Tidal}}\text{ .}
\end{equation}

The voltage difference between the two ends of a given dumbbell (which is a
model of the voltage difference between the opposite faces of a given cube)
is given by%
\begin{equation}
V=\beta \frac{Q}{4\pi \varepsilon _{0}L}\text{ ,}  \label{beta}
\end{equation}%
where $\beta $ is another pure number of the order of unity (again, this
follows from dimensional considerations, since $L$ is only distance scale in
the problem). Substituting the squared quantity $Q^{2}/L^{2}$ obtained from (%
\ref{beta}) into (\ref{alpha}), one gets%
\begin{eqnarray}
\alpha \frac{Q^{2}}{4\pi \varepsilon _{0}L^{2}} &=&\alpha \frac{(4\pi
\varepsilon _{0})^{2}V^{2}}{(4\pi \varepsilon _{0})\beta ^{2}}=4\pi
\varepsilon _{0}\frac{\alpha }{\beta ^{2}}V^{2}  \nonumber \\
&=&Mg^{\prime }\approx \rho L^{3}g\theta \approx \rho g\frac{L^{4}}{R_{E}}%
\text{ .}
\end{eqnarray}%
Solving for the voltage difference $V$, one obtains%
\begin{equation}
V\approx \left( \frac{\beta ^{2}}{\alpha }\frac{\rho gL^{4}}{4\pi
\varepsilon _{0}R_{E}}\right) ^{1/2}=\frac{\left\vert \beta \right\vert }{%
\sqrt{\alpha }}V_{\text{F-F}}\text{ ,}
\end{equation}%
where the characteristic free-fall voltage scale $V_{\text{F-F}}$ for
characteristic experimental parameters ($L=1$ cm, $\rho =10^{4}$ kg/m$^{3}$)
is given by%
\begin{equation}
V_{\text{F-F}}=\left( \frac{\rho gL^{4}}{4\pi \varepsilon _{0}R_{E}}\right)
^{1/2}\sim 1\text{ Volt ,}
\end{equation}%
which is experimentally interesting, given the reasonably large capacitances
of the two-cube system. (For the geometry of the dumbbells indicated in
Figure \ref{Ball-and-stick-model}, the numerical values of $\alpha =11/18$
and $\beta =-2/3$ are, as indicated earlier, on the order of unity.)

\section{The \textquotedblleft Cavendish-like\textquotedblright\ experiment}

The order-of-magnitude estimate given above indicates that experiments are
feasible. An experiment was performed at Merced, in which a slowly time-varying, inhomogeneous, tidal
gravitational field is produced by means of two piles of lead bricks placed
diametrically opposite each other on a slowly rotating, circular platform,
as the sources of the field. The two piles of bricks, which weigh
approximately a ton, will orbit slowly and symmetrically around a
superconducting circuit similar to the one shown in Figure 3, which is suspended inside a dilution
refrigerator above the center of the rotating platform.

\begin{figure}[tbp]\begin{center}
\label{pendula} 
\includegraphics[width=1.5in]{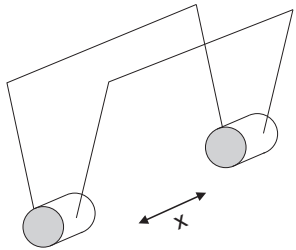}\end{center}
\caption{Pendula are supported by two wires each. \ This configuration causes motion along a single axis, denoted by $x$ in the figure.}
\end{figure}

For simplicity, we consider each pendulum to have two support wires, as shown in Figure \ref{pendula}, so that
the motion is constrained along one dimension.

\bigskip

The source masses will rotate around the dilution refrigerator containing
the pendula at approximately 1 rpm, and so the angular frequency is
approximately $\omega =\frac{\pi }{30}$ $s^{-1}.$ \ The gravitational forces
on the pendula due to the source masses will thus be periodic in time with a
period of approximately 60 seconds (a 30-second period is theoretically
possible due to symmetry, but for systematic errors in the setup and design,
we will assume 60 seconds). \ The magnitude of the gravitational forces will
vary according to Newton's inverse-square law. \ In each instant in time,
the forces on the pendula can be calculated. \ Assuming that the angular
frequency of the rotating source masses is sufficiently small, a
quasi-static model can be implemented, in which the system is in static
equilibrium at any given time.

Figure 6 shows the system at an arbitrary instant in time where $\theta
=\omega t$. \ The forces $\mathbf{F}_{1}$ and $\mathbf{F}_{2}$ are
gravitational forces exerted by the source masses. \ $\mathbf{F}_{3}$ is the
gravitational force exerted by the opposite pendulum. \ The separation
distance between the two pendula is $s$. \ Let the vertical faces of the
source masses that are (roughly) oriented radially have length $a$, the
vertical faces facing the dilution refrigerators have length $b$, and the
heights have dimension $c.$ 

\begin{figure}[ht]\begin{center}
\label{cav setup} 
\includegraphics[width=3in]{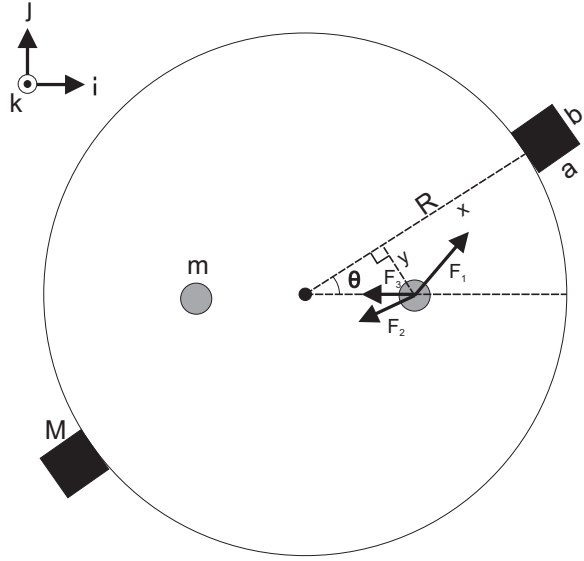}\end{center}
\caption{The experimental setup of the Cavendish-like experiment. \ The figure shows a snapshot of the dynamic system, where the small displacements of the pendula have, for the moment, been neglected. \ The coordinates $x,y$ and $z$ will vary with the angle $\protect\theta$, and $\mathbf{i, j}$ and $\mathbf{k}$ are unit basis vectors.}
\end{figure}

A differential mass element $dM$ of the upper source mass in Figure 6 exerts a differential force on the right pendulum with magnitude
(direction will be considered separately) given by%
\begin{equation}
\left\vert d\mathbf{F}_{1}\right\vert =\frac{Gm}{x^{2}+y^{2}+z^{2}}dM=\frac{%
Gm\rho }{x^{2}+y^{2}+z^{2}}dV,
\end{equation}%
where $\rho $ is the density of the material (assumed to be uniform). \
Integrating this expression, we obtain the total force exerted by the upper
source mass, given by%
\begin{equation}
\left\vert \mathbf{F}_{1}\right\vert =Gm\rho \int_{-\frac{c}{2}}^{\frac{c}{2}}\int_{\frac{s}{2}\sin
\omega t-\frac{b}{2}}^{\frac{s}{2}\sin \omega t+\frac{b}{2}}\int_{R-\frac{s}{2}\cos \omega t}^{R-\frac{s}{2}\cos
\omega t+a}\frac{dx\hspace{0.03in}dy \hspace{0.03in}dz}{x^{2}+y^{2}+z^{2}}.
\end{equation}%
Similarly, the total force exerted on the right pendulum by the lower source
mass in Figure 6 is given by%
\begin{equation}
\left\vert \mathbf{F}_{2}\right\vert =Gm\rho \int_{-\frac{c}{2}}^{\frac{c}{2}}\int_{\frac{s}{2}\sin
\omega t-\frac{b}{2}}^{\frac{s}{2}\sin \omega t+\frac{b}{2}}\int_{R+\frac{s}{2}\cos \omega t}^{R+\frac{s}{2}\cos
\omega t+a}\frac{dx\hspace{0.03in}dy \hspace{0.03in}dz}{x^{2}+y^{2}+z^{2}}.
\end{equation}

The angle $\beta $ subtended by vector $%
\mathbf{F}_{1}$ and the line joining the two pendula in Figure 6 satisfies the equations 
\begin{subequations}
\begin{eqnarray}
\cos \beta  &=&\sqrt{\frac{(R+\frac{a}{2})^{2}\cos ^{2}\omega t+\frac{s^{2}}{4}-s(R+\frac{a}{2})\cos
\omega t}{(R+\frac{a}{2})^{2}+\frac{s^{2}}{4}-s(R+\frac{a}{2})\cos \omega t}} \\
\sin \beta  &=&\frac{(R+\frac{a}{2})\sin \omega t}{\sqrt{%
(R+\frac{a}{2})^{2}+\frac{s^{2}}{4}-s(R+\frac{a}{2})\cos \omega t}}
\end{eqnarray}%
and the angle $\gamma $ subtended by $\mathbf{F}_{2}$ and the line
joining the two pendula in Figure 6 satisfies the equations 
\end{subequations}
\begin{subequations}
\begin{eqnarray}
\cos \gamma  &=&\frac{\frac{s}{2R+a}+\cos \omega t}{\sqrt{\left( \frac{s}{%
2R+a}+\cos \omega t\right) ^{2}+\sin ^{2}\omega t}}  \label{gamma} \\
\sin \gamma  &=&\frac{\sin \omega t}{\sqrt{\left( \frac{s}{2R+a}+\cos \omega
t\right) +\sin ^{2}\omega t}},
\end{eqnarray}%
and thus the full expressions for the forces acting on the right pendulum in
Figure 6 are given by 
\end{subequations}
\begin{eqnarray}
\mathbf{F}_{1} &=&Gm\rho \int_{-\frac{c}{2}}^{\frac{c}{2}}\int_{\frac{s}{2}\sin \omega
t-\frac{b}{2}}^{\frac{s}{2}\sin \omega t+\frac{b}{2}}\int_{R-\frac{s}{2}\cos \omega t}^{R-\frac{s}{2}\cos \omega t+a}%
\frac{dx\hspace{0.03in}dy \hspace{0.03in}dz}{x^{2}+y^{2}+z^{2}}\left( \cos \beta \text{ }\mathbf{i}+\sin
\beta \text{ }\mathbf{j}\right)  \\
\mathbf{F}_{2} &=&-Gm\rho \int_{-\frac{c}{2}}^{\frac{c}{2}}\int_{\frac{s}{2}\sin \omega
t-\frac{b}{2}}^{\frac{s}{2}\sin \omega t+\frac{b}{2}}\int_{R+\frac{s}{2}\cos \omega t}^{R+\frac{s}{2}\cos \omega t+a}%
\frac{dx\hspace{0.03in}dy \hspace{0.03in}dz}{x^{2}+y^{2}+z^{2}}\left(\cos \gamma \text{ }\mathbf{i}+\sin
\gamma \text{ }\mathbf{j}\right) ,
\end{eqnarray}%
assuming that the pendula and the centers of mass of the source masses are
coplanar. \ Recalling that the pendula are constrained to move along the
axis parallel to the unit vector $\mathbf{i}$, the unconstrained components
of the forces exerted by the source masses on the right pendulum are given
by 
\begin{eqnarray}
F_{1} =Gm\rho \cos \beta \int_{-\frac{c}{2}}^{\frac{c}{2}}\int_{\frac{s}{2}\sin \omega
t-\frac{b}{2}}^{\frac{s}{2}\sin \omega t+\frac{b}{2}}\int_{R-\frac{s}{2}\cos \omega t}^{R-\frac{s}{2}\cos \omega t+a}%
\frac{dx\hspace{0.03in}dy \hspace{0.03in}dz}{x^{2}+y^{2}+z^{2}} \\
F_{2} =Gm\rho \cos \gamma \int_{-\frac{c}{2}}^{\frac{c}{2}}\int_{\frac{s}{2}\sin \omega
t-\frac{b}{2}}^{\frac{s}{2}\sin \omega t+\frac{b}{2}}\int_{R+\frac{s}{2}\cos \omega t}^{R+\frac{s}{2}\cos \omega t+a}%
\frac{dx\hspace{0.03in}dy \hspace{0.03in}dz}{x^{2}+y^{2}+z^{2}}.
\end{eqnarray}

\begin{figure}[t]\begin{center}
\label{displacement} 
\includegraphics[width=2in]{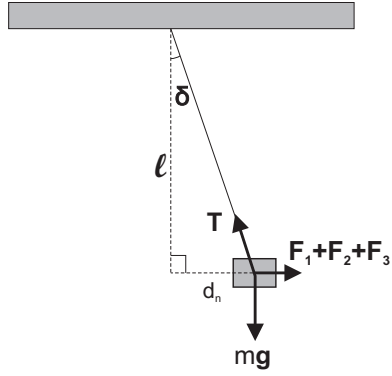}\end{center}
\caption{Force body diagram of single pendulum.}
\end{figure}

The force exerted by the left pendulum on the right pendulum is
anti-parallel to $\mathbf{i}$, and its magnitude is given by 
\begin{equation}
F_{3}=\frac{Gm^{2}}{s^{2}}.
\end{equation}

Viewing the right pendulum from the side, the free body diagram can be
pictured as in Figure 7.

Comparing the vertical
and horizontal components of the vectors in Figure 7, we
obtain an expression for the deflection magnitude $d_{n}$, given by
\begin{equation}
d_{n}\approx \frac{\ell }{mg}\left( F_{1}-F_{2}-F_{3}\right) ,
\label{deflection}
\end{equation}%
where we have assumed $d_{n}$ to be small, so that (\ref{deflection}) is
explicit. \ The subscript $n$ is used for the case that the pendula are
normal, so that both the ionic lattice and the valence electron system
follow the same local geodesics. \ $F_{3}$ can be safely
neglected if the source masses are sufficiently large.

\bigskip

Let us now turn to the concept of charge separation within superconductors.
\ Let the pendula now be superconducting, and share a continuous
superconducting connection, so that the Cooper-pair wavefunction has a
constant phase across both pendula. \ If we assume that only the ionic
lattice of the pendula are subject to the gravitational forces of the source
masses, a charge separation will ensue. \ For the moment, let us assume that
the Cooper pair wavefunction is completely rigid, and that the superfluid
will therefore remain motionless with respect to the center of mass of the
pendulum system, which lies approximately at the midpoint between the two
pendula. \ This charge separation will cause Coulomb forces both between the
two pendula, and within the pendula themselves. \ These forces must be taken
into account to accurately predict the motions of the ionic lattices, and
the magnitude of the charge separation.

\begin{figure}[t]\begin{center}
\label{ball & stick} 
\includegraphics[width=3in]{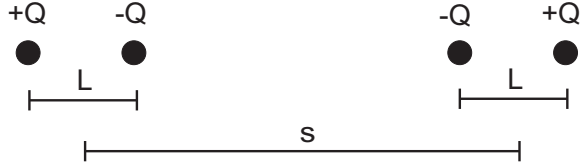}\end{center}
\caption{Superconducting pendula under the influence of tidal forces that produce charge sepration. \ The Physical dimension of the pendulum along the axis of the charge separation is given by $L$, and the center-to-center distance between the pendula is given by $s$.}
\end{figure}
Figure 8 shows the pendula at an instant in time when the
ionic lattices are pulled outwards by the source masses. \ The charged faces
of the pendula are modeled here using point charges.

Summing the Coulombic forces that the left pendulum exerts on the ionic
lattice of the right pendulum in Figure 8, it can be shown
that the total Coulombic force is%
\begin{equation}
\mathbf{F}_{Q,d>0}\approx \frac{-Q^{2}}{4\pi \epsilon _{0}s^{2}}\left[ \frac{%
L(2s+L)}{(s+L)^{2}}\right] \mathbf{i,}  \label{Coul_force_ext_d>0}
\end{equation}%
where we have ignored the charge extrusion distance $d$, since $d$ is much
smaller than $s$ and $L$. \ The Coulomb force on the ionic lattice of the
right pendulum exerted by the electron superfluid within the right pendulum,
due to charge separation, is%
\begin{equation}
\mathbf{F}_{c,d>0}\approx -\frac{Q^{2}}{4\pi \epsilon _{0}L^{2}}\mathbf{i},
\label{Coul_force_int_d>0}
\end{equation}%
where the same approximation has been used. \ The subscript $d>0$ has been
used since (\ref{Coul_force_ext_d>0}) and (\ref{Coul_force_int_d>0}) are
only valid when the positively-charged ionic lattice is extruded from the
negatively-charged superfluid on the outer faces of the pendula. \ When the
signs of the charges are reversed, the forces on the ionic lattice of the
right pendulum can be found by making the substitution $L\rightarrow -L$, so
the general expression for the Coulombic forces on the right pendulum are%
\begin{equation}
\mathbf{F}_{Q}\approx \frac{-Q^{2}}{4\pi \epsilon _{0}s^{2}}\left\{ \frac{L%
\text{sgn}(d)\left[ 2s+L\text{sgn}(d)\right] }{\left[ s+L\text{sgn}(d)\right]
^{2}}\right\} \mathbf{i}  \label{Coul_force_ext}
\end{equation}%
and%
\begin{equation}
\mathbf{F}_{c}\approx -\frac{Q^{2}\text{sgn}(d)}{4\pi \epsilon _{0}L^{2}}%
\mathbf{i},  \label{Coul_force_int}
\end{equation}%
where%
\begin{equation}
\text{sgn}(d)=\left\{ 
\begin{array}{c}
1,\text{ }d>0 \\ 
0,\text{ \ }d=0 \\ 
-1,\text{ \ }d<0%
\end{array}%
\right. 
\end{equation}

Considering all forces on a single superconducting pendulum, we have a model
that is similar to that depicted in Figure 7, but $\mathbf{F%
}_{Q}$ and $\mathbf{F}_{c}$ are considered, in addition to $%
\sum_{i=1}^{3}\mathbf{F}_{i}.$ For this case, we have%
\begin{equation}
d\approx \frac{\ell }{mg}\left( F_{1}-F_{2}-F_{3}-F_{Q}-F_{c}\right) ,
\label{d1}
\end{equation}%
where $F_{Q}$ and $F_{c}$ are the magnitudes of the vectors defined in (\ref%
{Coul_force_ext}) and (\ref{Coul_force_int}), respectively.

The extruded charge $Q$ depends on the extrusion length. \ Starting with%
\begin{equation}
Q=2en_{s}V,  \label{charge}
\end{equation}%
where $n_{s}$ is the superconducting electron density, $-2e$ is the Cooper
pair charge, and $V$ is the extrusion volume. \ The current experimental
design involves cylindrically-shaped pendula, so (\ref{charge}) becomes%
\begin{equation}
Q=2en_{s}\pi r^{2}d,
\end{equation}%
where $r$ is the cylindrical radius of the sample. \ Thus,%
\begin{equation}
d=\frac{Q}{2en_{s}\pi r^{2}},  \label{d2}
\end{equation}%
and it should be noted here that%
\begin{equation}
\text{sgn}(d)=\text{sgn}(Q).  \label{sgn}
\end{equation}%
Substitution of (\ref{d2}) and (\ref{sgn}) into (\ref{d1}), we have%
\begin{equation}
Q=\frac{2en_{s}\pi r^{2}\ell }{mg}\left( F_{1}-F_{2}-F_{3}-\frac{Q^{2}\text{%
sgn}(Q)}{4\pi \epsilon _{0}s^{2}}\left\{ \frac{L\left[ 2s+L\text{sgn}(Q)%
\right] }{\left[ s+L\text{sgn}(Q)\right] ^{2}}\right\} -\frac{Q^{2}\text{sgn}%
(Q)}{4\pi \epsilon _{0}L^{2}}\right) .  \label{charge_exp}
\end{equation}%
An iterative numerical method was used to solve (\ref{charge_exp}) for $Q$.  \ It should also be noted that the expressions for $F_{1}$
and $F_{2}$ were evaluated using numerical integration.  A plot of the expected charge as a function of time using actual experimental parameters is depicted in Figure 9.

\bigskip

\begin{figure}[tbp] \begin{center}
\includegraphics[width=2.5in]{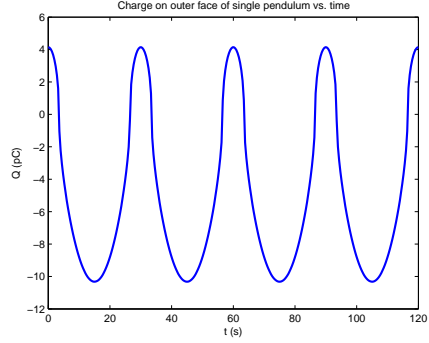}\end{center}
\caption{Predicted charge signal as a function of time.}
\end{figure}

We expect to be able to see (using synchronous detection) the charge
separation induced by these gravitational fields in a superconducting
circuit, which consists of two well-separated superconducting bodies, both
of which are suspended by means of pairs of superconducting wires inside the
same refrigerator, so that the two bodies form the superconducting plumb
bobs of two pendula. These two bodies are then \emph{coherently} connected
to each other by means of a pair of parallel superconducting wires, as
indicated in Figure 3, to form a single
superconducting circuit, i.e., a single quantum entity. The charge
separation effect can then be measured inductively by means of a sensitive
electrometer (we should be able to see the charge induced by the extruded
Cooper pairs, which should be on the order of picocoulombs, with a high
signal-to-noise ratio). If we should observe a nonzero charge-separation
signal in this experiment, then this observation would establish the
existence of a novel coupling between gravity and electricity mediated by
means of macroscopic quantum matter.

Normally, the gravitational fields due to the lead bricks should cause small
angular deflections on the order of nanoradians of the two pendula. These
small deflections should occur relative to the local vertical axis which is
located at the midpoint in between them (see Figure \ref%
{radial-converging-geodesics}, in which the two freely-falling objects are
replaced by the two plumb bobs of the two pendula). We would normally expect
to see such deflections if these two pendula consisted of normal, classical
matter, or if they consist of two superconducting plumb bobs which have had
their superconducting connection between them destroyed due to decoherence.
Such deflections could be measured with high signal-to-noise ratios using
laser interferometry. If we were to monitor both the deflections of the
pendula and the charge separation effect in the same experiment, there would
be four logical possibilities as to the possible outcomes:

(I) Charge-separation? YES. Deflection? NO.

(II) Charge-separation? NO. Deflection? YES.

(III) Charge-separation? YES. Deflection? YES.

(IV) Charge-separation? NO. Deflection? NO.

Based on the arguments presented above, we would expect (I) to be the
outcome, if the Cooper pairs were to be able to drag the ions of the lattice
into co-motion with these superconducting electrons during free fall. In the
tug-of-war between the uncertainty principle and the equivalence principle,
the uncertainty principle wins in (I). By contrast, if there is nothing
special about this superconducting system over any other material system,
i.e., if the universality of free fall were to apply to the Cooper pairs
inside the superconducting circuit so that they would undergo free fall
along with the ions, and therefore the superconducting system would remain
electrically neutral and unpolarized during free fall, then we would expect
(II) to be the outcome. The equivalence principle wins in (II). If, however,
there does exist a charge-separation effect, but the ions of the lattice
were to drag the Cooper pairs into co-motion with the ionic lattice during
free fall, then we would expect (III) to be the outcome. Finally, there
exists the remote possibility of outcome (IV), which would indicate that
Newtonian gravity would somehow have failed to produce any deflection at all
of the pendula in this experiment, but that nevertheless the system remains
electrically neutral and unpolarized in the presence of the ton of bricks.
Results from this \textquotedblleft Cavendish-like\textquotedblright\
experiment will be presented elsewhere.

Under the present experimental conditions of the Cavendish-Like experiment,
two of the four outcomes listed above have been ruled out. To understand why
let us consider one of the superconducting pendula plumb bobs, and suppose
that a stable DC charge separation effect exist within the plumb bob for $%
t\leq 0$. Note that, as shown above, this DC charge separation effect would
lead to an internal voltage drop of approximately 1 Volt across the
superconducting plumb bob. By stable, it is meant that there are no time
changing quantities so that 
\[
\frac{\partial \Phi _{B}}{\partial t}=0,~\frac{\partial A}{\partial t}=0~%
\text{and}~\frac{\partial \Phi _{E}}{\partial t}=0;~~t\leq 0.
\]%
Because there is no external electric field to cancel out the internal
electric field produced by the predicted stable DC charge separation effect
for $t\leq 0$, there must be a measurable voltage drop across the
superconducting plumb bob. This is to be contrasted with the case of a
conductor placed between two parallel plates of a capacitor where there
exists a stable charge separation effect but no measurable voltage drop
across it. Hence, if we allow the system to evolve at $t>0$, the Cooper
pairs deep inside the SC will experience a scalar potential ($\phi $) which
imparts energy to the pairs. 
\[
|E|=q\phi =2e\phi \sim 2~\text{eV},
\]%
where we set $\phi =1$ Volt as predicted above. Comparing this energy with
the ground state energy gap involved in most conventional type I
superconductors, $E_{g}\sim \text{meV}$, we see that such scalar potential
would excite the Cooper pairs into the quasi-particle state, thus leading to
a pair-breaking mechanism. Therefore, such DC charge separation effect would
not be stable and there would be no measurable charge separation in the
present Cavendish-Like experiment. Hence outcomes (I) and (III), where a
measurable stable DC charge separation effect are expected, have already
been ruled out, leaving us with outcome (II) and (IV) as the only remaining
two possibilities, though in an experiment involving high-frequency perturbations, these two outcome possibilities may remain.

\section{Acknowledgements}

The authors thank Prof. Michael Scheibner, Bong-Soo Kang, and Philip Jensen
for their assistance concerning the Cavendish-like experiment.

Raymond Chiao wishes to extend his heartiest birthday congratulations to
Danny and Helmut!

\end{document}